\documentclass[showpacs,amsmath,amssymb,twocolumn,prl,superscriptaddress]{revtex4-2}
\usepackage{amssymb}
\usepackage[dvips]{graphicx}
\usepackage{enumerate}
\usepackage{epsfig} 
\usepackage{subfigure}
\usepackage{xcolor}
\usepackage[T1]{fontenc}
\usepackage{fullpage}
\usepackage{amsthm,amsfonts,amssymb,amscd,mathrsfs,xspace,framed}
\usepackage{mathrsfs,amsmath}
\usepackage{color}
\usepackage{setspace}
\usepackage{url}
\usepackage{wrapfig}
\usepackage{tikz}
\usepackage{enumitem}
\usepackage{bm}
\usepackage{comment}
\usepackage{txfonts}
\usepackage{array}
\usepackage{color}
\usepackage{braket}
\usepackage{natbib}
\usepackage{cleveref}
\usepackage{graphicx}
\usepackage{pdfpages}

\def\be{\begin{equation}}
\def\ee{\end{equation}}
\def\ba{\begin{eqnarray}}
\def\ea{\end{eqnarray}}

\makeatletter
\AtBeginDocument{\let\LS@rot\@undefined}
\makeatother

\def\be{\begin{equation}}
\def\ee{\end{equation}}
\def\ba{\begin{eqnarray}}
\def\ea{\end{eqnarray}}

\begin{document}

\title{Demonstration of Entanglement-Enhanced Covert Sensing}

\author{Shuhong Hao}
\affiliation{
Department of Materials Science and Engineering, University of Arizona, Tucson, Arizona 85721, USA
}

\author{Haowei Shi}
\affiliation{
James C. Wyant College of Optical Sciences, University of Arizona, Tucson, Arizona 85721, USA
}

\author{Christos N. Gagatsos}
\affiliation{
James C. Wyant College of Optical Sciences, University of Arizona, Tucson, Arizona 85721, USA
}

\author{Mayank Mishra}
\affiliation{
James C. Wyant College of Optical Sciences, University of Arizona, Tucson, Arizona 85721, USA
}

\author{Boulat Bash}
\affiliation{
Department of Electrical and Computer Engineering, University of Arizona, Tucson, Arizona 85721, USA
}
\affiliation{
James C. Wyant College of Optical Sciences, University of Arizona, Tucson, Arizona 85721, USA
}

\author{Ivan Djordjevic}
\affiliation{
Department of Electrical and Computer Engineering, University of Arizona, Tucson, Arizona 85721, USA
}
\affiliation{
James C. Wyant College of Optical Sciences, University of Arizona, Tucson, Arizona 85721, USA
}

\author{Saikat Guha}
\affiliation{
James C. Wyant College of Optical Sciences, University of Arizona, Tucson, Arizona 85721, USA
}
\affiliation{
Department of Electrical and Computer Engineering, University of Arizona, Tucson, Arizona 85721, USA
}

\author{Quntao Zhuang}
\affiliation{
Department of Electrical and Computer Engineering, University of Arizona, Tucson, Arizona 85721, USA
}
\affiliation{
James C. Wyant College of Optical Sciences, University of Arizona, Tucson, Arizona 85721, USA
}

\author{Zheshen Zhang}
\email{zsz@arizona.edu}
\affiliation{
Department of Materials Science and Engineering, University of Arizona, Tucson, Arizona 85721, USA
}
\affiliation{
Department of Electrical and Computer Engineering, University of Arizona, Tucson, Arizona 85721, USA
}
\affiliation{
James C. Wyant College of Optical Sciences, University of Arizona, Tucson, Arizona 85721, USA
}

\begin{abstract}
The laws of quantum physics endow superior performance and security for information processing: quantum sensing harnesses nonclassical resources to enable measurement precision unmatched by classical sensing, whereas quantum cryptography aims to unconditionally protect the secrecy of the processed information. Here, we present the theory and experiment for entanglement-enhanced covert sensing, a paradigm that simultaneously offers high measurement precision and data integrity by concealing the probe signal in an ambient noise background so that the execution of the protocol is undetectable with a high probability. We show that entanglement offers a performance boost in estimating the imparted phase by a probed object, as compared to a classical protocol at the same covertness level. The implemented entanglement-enhanced covert sensing protocol operates close to the fundamental quantum limit by virtue of its near-optimum entanglement source and quantum receiver. Our work is expected to create ample opportunities for quantum information processing at unprecedented security and performance levels.

\end{abstract}

\maketitle

{\em Introduction.---}Quantum information processing (QIP) hinges on nonclassical effects such as superposition and entanglement to enable new communication~\cite{bennett1999entanglement,ekert1991quantum,shi2020,hao2021entanglement}, sensing~\cite{giovannetti2006quantum,giovannetti2011advances,ligo2011gravitational,aasi2013enhanced,tse2019quantum,xia2020demonstration,pirandola18quantumsensing}, and computing~\cite{shor1999polynomial,biamonte2017quantum} capabilities beyond the reach of classical physics. Among these, quantum cryptography~\cite{pirandola2020advances,xu2020secure,djordjevic19QKD} has been envisaged to shift the landscape of information security and has migrated from proof-of-concept demonstrations in laboratory settings~\cite{pirandola2020advances,xu2020secure,djordjevic19QKD,lodewyck2007quantum,korzh2015provably,zhang2017floodlight,zhang2018experimental} to real-world intercontinental links relayed by a satellite~\cite{liao2017satellite,liao2018satellite,yin2017satellite}. Quantum cryptography protocols have now been embodied in a variety of realms, including blind quantum computing~\cite{barz2012demonstration}, decision making~\cite{ben2005fast}, and information gathering~\cite{ganz2017quantum}, to safeguard information from being acquired by an adversary. 

Quantum covert protocols have recently emerged to offer a feature beyond the scope of these quantum cryptography protocols---the executions of the very protocols, with a high probability, are undetectable from the adversary's perspective~\cite{bash2015quantum,azadeh16quantumcovert-isit,tahmasbi2020covert,tahmasbi2020toward,bullock2019fundamental,gagatsos20codingcovcomm,bash12sqrtlawisit,bash2013limits,bash2015hiding}, thereby ensuring the data integrity. The covertness of these protocols is fundamentally guaranteed by the indistinguishability between quantum states and hence can be quantified by the quantum measurement theory. In analogy to many quantum cryptography protocols~\cite{xu2015experimental,lo2012measurement,lo2005decoy}, quantum covert protocols may be solely constructed upon classical transmitters and receivers~\cite{bash2015quantum,azadeh16quantumcovert-isit,bash17qcovertsensingisit,goeckel17sensinglinsystems-asilomar,gagatsos2019covert,gagatsos20codingcovcomm,bullock2019fundamental,gagatsos20codingcovcomm}, but quintessential quantum resources such as entanglement may offer additional performance gains. Indeed, the benefit of entanglement in quantum covert protocols has been recently analyzed~\cite{tahmasbi2021covert,shi2020,gagatsos20codingcovcomm}, but an experimental realization for entanglement-enhanced covert systems remains elusive. 

In this paper, we propose and experimentally implement an {\em entanglement-enhanced} covert sensing protocol and benchmark its performance against covert sensing based on classical resources~\footnote{In the remainder of the paper we will use the term ``classical covert sensing'' for covert sensing based on classical resources, even though the security of such schemes is guaranteed by quantum mechanics.} first presented in Refs.~\cite{bash17qcovertsensingisit,gagatsos2019covert}.
Both protocols are proven quantum-optimum in their own classes, and our experiment demonstrates that an entanglement transmitter, in conjunction with a quantum receiver, enables a 46.5\% reduction of the mean squared error (MSE) in estimating the phase imparted by an interrogated object, corresponding to a 87.6\% signal-to-noise ratio (SNR) improvement. Remarkably, the entanglement-enhanced covert sensing experiment operates at only 10\% off the ultimate quantum limit for the MSE. This work would spark new QIP applications fueled by entanglement-enhanced security and performance.

\begin{figure}[bth]
    \centering
    \includegraphics[width=0.45\textwidth]{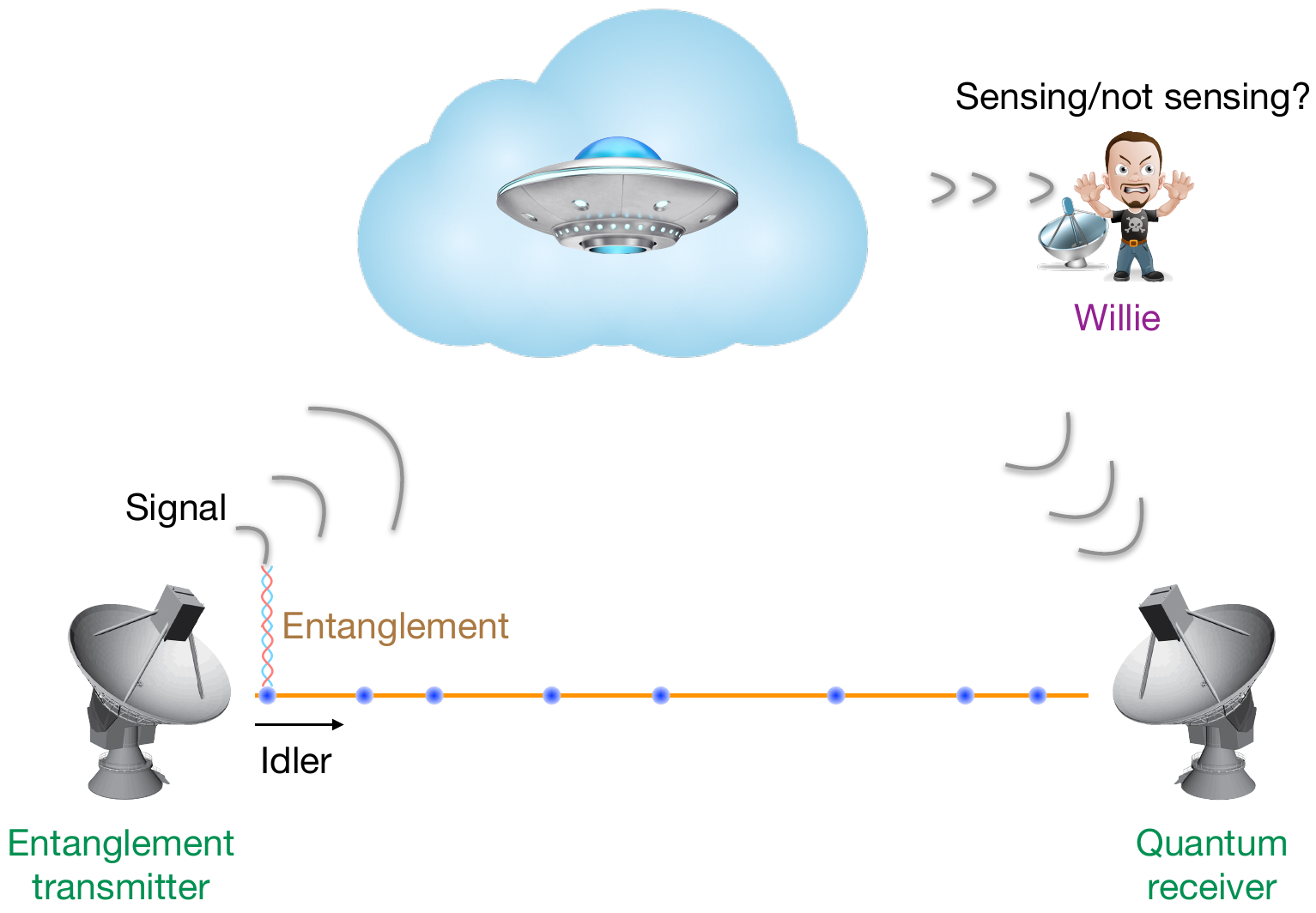}
    \caption{Configuration for entanglement-enhanced covert sensing. Entanglement transmitter generates entangled signal and idler and sends the signal to probe an object. The quantum receiver performs a joint measurement on the signal returned from a lossy and noisy environment and the locally stored idler. Willie takes the optimal quantum measurement to detect the sensing attempt.}
    \label{fig:Concept}
\end{figure}

{\em Protocols.---}Sketched in Fig.~\ref{fig:Concept}, the covert sensor comprises an entanglement transmitter and a quantum receiver, aimed at probing the phase shift imparted by an object situated in a lossy and noisy environment characterized by the overall transmissivity $\kappa_E$ and the average per-mode background-noise photon number $N_B$. In covert sensing, the transmitter prepares $M$ copies of entangled signal-idler mode pairs, represented as $\hat{\rho}_{SI}^{\otimes M}$, with on average $N_S$ photons per mode. The idler modes are locally retained in a quantum memory with efficiency $\kappa_I$. The transmissivity for the signal modes within the entanglement transmitter is $\kappa_T$. The signal modes are exploited to interrogate a phase shift $\theta$ imparted by an object, yielding the global state $\hat{\rho}_{SI}^{'\otimes M}(\theta)$. Accounting for the channel loss and environmental background noise, each signal mode at the quantum receiver carries on average $N_B$ noise photons. The overall transmissivity between the entanglement source and the quantum receiver is defined as $\kappa \equiv \kappa_T\kappa_E$. The quantum receiver takes a joint measurement on the signal-idler mode pairs to generate an estimator for the phase: $\mathcal{M}_Q\left[\hat{\rho}_{SI}^{'\otimes M}(\theta)\right] \rightarrow \hat{\theta}_Q$. The estimation precision is quantified by the root-mean-square (rms) error $\delta \theta_Q = \sqrt{\left\langle\left(\theta-\hat{\theta}_Q\right)^2\right\rangle}$ subject to the quantum Cram\'{e}r-Rao bound (QCRB):
\begin{equation}
    \delta \theta_Q^2 \geq \frac{1}{M\mathcal{J}},
\end{equation}
where $\mathcal{J}$ is the quantum Fisher information (QFI) for the sensing protocol under investigation (see Ref.~\cite{supp}).

\begin{figure*}[bth]
    \centering
    \includegraphics[width=1\textwidth]{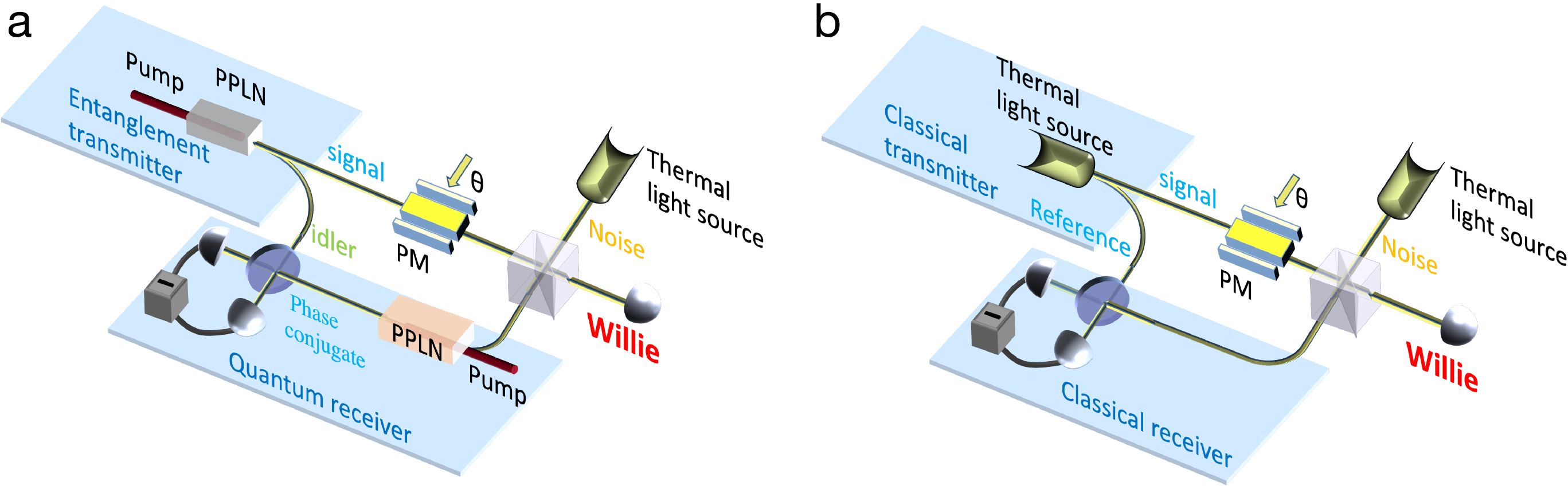}
    \caption{Experimental setups for (a) entanglement-enhanced covert sensing and (b) covert sensing based on thermal light. Willie's apparatus to detect the sensing attempt is illustrated in Ref.~\cite{supp}. PPLN: periodically poled lithium niobate. PM: phase modulator.}
    \label{fig:Setup_maintext}
\end{figure*}

To detect the sensing attempt, the adversary, Willie, captures the probe photons lost in the noisy environment and endeavors to discriminate between two quantum states, $\hat{\rho}_0^{\otimes M}$ for the sole background noise and $\hat{\rho}_1^{\otimes M}$ for the same background noise augmented by a weak probe signal. The lower bound of Willie's detection error probability under such a quantum-state discrimination problem is given by 
\begin{equation}
    \mathbb{P}_e^{\rm (w)} \geq \frac{1}{2} \left(1-\frac{1}{2}\left\Vert\hat{\rho}_0^{\otimes M}- \hat{\rho}_1^{\otimes M}\right\Vert_1\right) \geq \frac{1}{2} - \epsilon,
\end{equation}
where $\left\Vert\hat{\rho}_0^{\otimes M}- \hat{\rho}_1^{\otimes M}\right\Vert_1$ is the trace distance between $\hat{\rho}_0^{\otimes M}$ and $\hat{\rho}_1^{\otimes M}$, and $\epsilon$ is the covertness parameter, which can be arbitrarily set by choosing $N_S$ and $M$. Within the range for the operational parameters of interest~\cite{supp}, 
\begin{equation}
\label{eq:epsilon}
   \epsilon \propto \frac{\sqrt{M}N_S}{N_B}.
\end{equation}

The entanglement-enhanced covert sensing protocol is benchmarked against covert sensing based on classical states to demonstrate a quantum advantage. In the classical protocol, the sensor employs $M$ copies of the probe state, $\hat{\rho}_S^{\otimes M}$ with the same energy as the entanglement-enhanced case, to interrogate the same phase object, resulting in $\hat{\rho}_S^{'\otimes M}(\theta)$ at the quantum receiver. A measurement $\mathcal{M}_C\left[\hat{\rho}_S^{'\otimes M}(\theta)\right]$ then produces a phase estimator $\hat{\theta}_C$ with the rms error $\delta \theta_C$. The marginal states $\hat{\rho}_0^{\otimes M}$ and $\hat{\rho}_1^{\otimes M}$ for Willie are set identical in the entanglement-enhanced and classical protocols so that their performance levels are evaluated under the same covertness parameter.

{\em Experiment.---}Our experimental setup is illustrated in Fig.~\ref{fig:Setup_maintext}a, with a detailed description enclosed in Ref.~\cite{supp}. The transmitter consists of a periodically poled lithium niobate (PPLN) crystal to generate non-degenerate entangled signal and idler modes each occupying an optical bandwidth of $W$. The signal photons are exploited to probe a phase shift $\theta$ induced by an phase modulator (PM) while the idler photons are locally stored in a spool of low-loss optical fibers. Sensing is executed over $T$ seconds consuming $M= WT$ signal-idler mode pairs. The environmental noise is emulated by injecting thermal noise from an amplified spontaneous emission (ASE) source through a beam splitter. To infer the phase shift, a joint measurement is performed on the returned noisy signal and the retained idler modes in a phase-conjugate receiver (PCR)~\cite{guha2009gaussian}, which has been employed in entanglement-assisted communication to surpass the ultimate classical capacity~\cite{hao2021entanglement}. In the PCR, the returned signal and the pump are combined at a second PPLN crystal to produce phase-conjugate modes via a low-gain difference-frequency generation process, through which the phase-sensitive cross correlation between the signal and idler modes is carried over to the phase-insensitive cross correlation between the phase-conjugate and idler modes, while only a small amount of noise in the signal modes is converted to the phase-conjugate modes. The wavelength of the phase-conjugate modes matches that of the idler modes, allowing them to interfere on a 50:50 beam splitter. The two output arms of the beam splitter are measured by a pair of photodetectors in a balanced setting to produce difference photocurrent, from which the phase estimator $\hat{\theta}_Q$ is acquired. The quantum advantage reaped by the PCR stems from the initial phase-sensitive cross correlation between the entangled signal and idler modes. The residue phase-sensitive cross correlation utilized by the PCR, albeit substantially weakened by the environment, remains much stronger than any classical probe and reference can deliver.

We also build a covert-sensing setup with classical resources as a performance benchmark (Fig.~\ref{fig:Setup_maintext}b). In that experiment, the output of a thermal-light source is split into a signal arm and a reference arm. Compared to a coherent-light source, the thermal-light source features a large optical bandwidth proven advantageous for covert sensing~\cite{gagatsos2019covert}. The signal photons are modulated by the PM. At the homodyne receiver (HR), the returned signal photons mix with the reference on a 50:50 beamsplitter followed by two photodetectors to take a balanced measurement that constructs the classical phase estimator $\hat{\theta}_C$. 

To detect the sensing attempt, Willie takes a measurement in the noise background on a portion of the signal photons. Since Willie's marginal states $\hat{\rho}_0$ and $\hat{\rho}_1$ are both thermal, direct photon counting on a photodetector constitutes his optimal measurement for this quantum-state discrimination task to infer the presence of the probe. Willie's error probability in detecting the sensing attempt is tested by repeating a series of such measurements taken with or without the probe signal.

\begin{figure}[bth]
    \centering
    \includegraphics[width=0.5\textwidth]{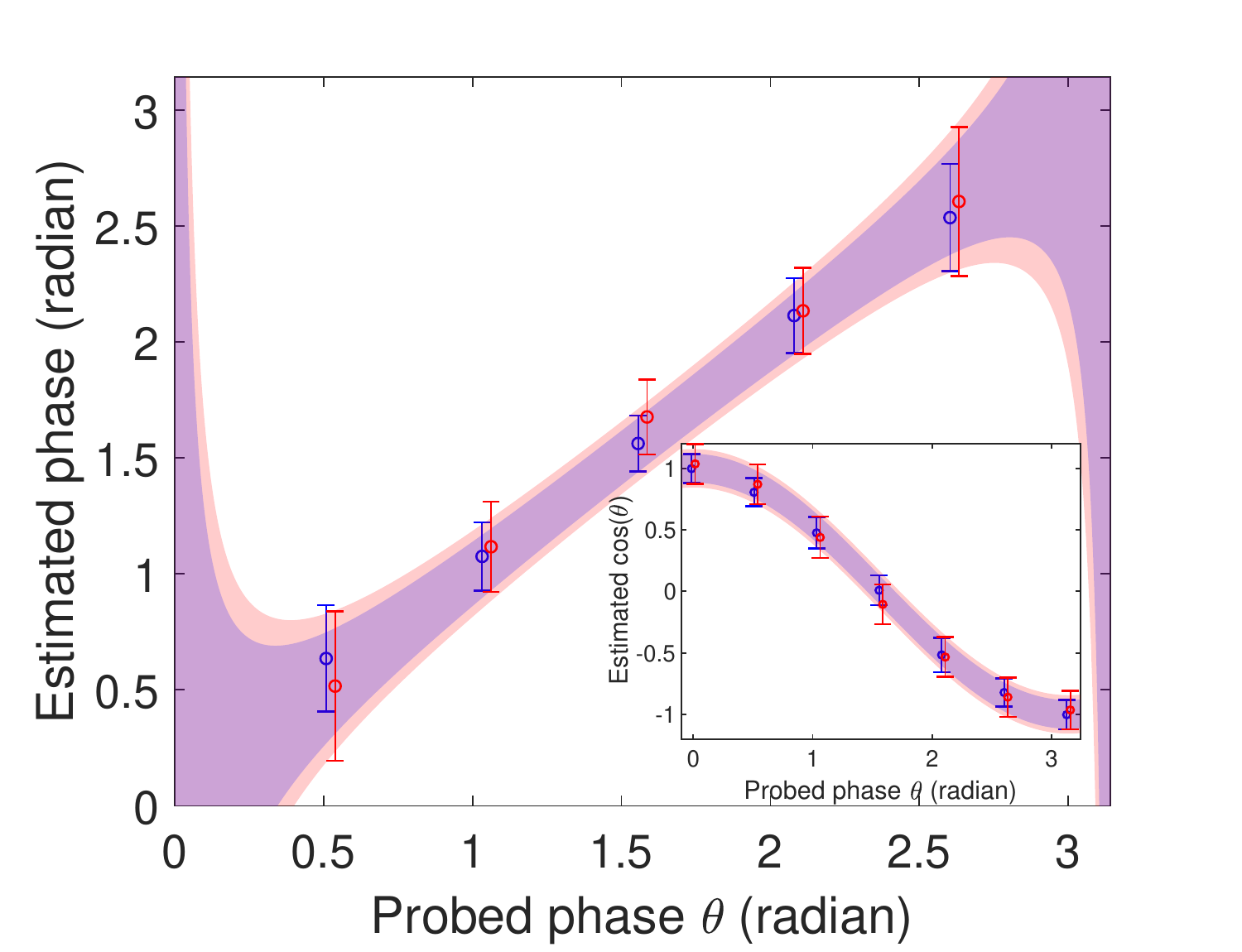}
    \caption{Phase estimation in entanglement-enhanced covert sensing (blue) and classical covert sensing (red). Dots are experiment data, each obtained from averaging over 2000 consecutive measurements. Error bars represent rms errors. Shades represent the theoretical rms. Inset: estimation of $\cos(\theta)$. The probed phase values for classical (entanglement-enhanced) covert sensing are shifted on abscissa by 0.015 (-0.015) for readability. $N_S=8\times 10^{-4}$, $N_B=160$, $T$ = 125 $\mu$s, $\kappa$=0.0165.}
    \label{fig:Phase_est}
\end{figure}

We first assess the performance of phase estimation in terms of the rms errors for both covert sensing protocols subject to the same covertness parameter $\epsilon$, achieved by setting the brightness of the probes identical. An electro-optic modulator applies test phase shifts $\theta \in [0,\pi]$ on the probe in either sensing scenario. With appropriate scaling factors, the output of the PCR and the HR yield, respectively, unbiased cosine estimators $\cos\left(\hat{\theta}_Q\right)$ and $\cos\left(\hat{\theta}_C\right)$, as plotted in the inset of Fig.~\ref{fig:Phase_est}. The experimental data show good agreement with the theoretical model, demonstrating a quantum advantage for entanglement-enhanced covert sensing, manifested as a reduced experimental (error bars) and theoretical (shaded areas) estimation rms errors. The experimentally measured cosine estimation rms error averaged over all test phases arrives at 0.1220 $\pm$ 0.0088 for entanglement-enhanced covert sensing, as compared to 0.1614 $\pm$ 0.0036 for classical covert sensing. The uncertainties in the rms errors account for the source-brightness fluctuation caused by the power instabilities of the pump laser ($< \pm 1\%$), the ASE source ($<\pm 1\%$), and the free-space to fiber coupling efficiency variation ($< \pm 3\%$), along with other optical, electrical, and mechanical instabilities. To derive the phase estimators, we take the inverse function on the cosine estimators to acquire $\hat{\theta}_Q = \arccos\left[\cos(\hat{\theta}_Q)\right]$ and $\hat{\theta}_C = \arccos\left[\cos(\hat{\theta}_C)\right]$. Figure~\ref{fig:Phase_est} depicts the estimated phases vs the applied phases, showing that the rms error of the phase estimator for entanglement-enhanced covert sensing is reduced by an average of 24.0 \% from that of classical covert sensing.

We next study the performance of covert sensing under two environmental conditions at increasing background noise levels: the fixed covertness regime in which the probe power is adjusted to render the estimation MSE and Willie's detection error probability unchanged; and the fixed probe power regime in which the covertness is enhanced at the cost of an increased estimation MSE.

\begin{figure}[bth]
    \centering
    \includegraphics[width=0.5\textwidth]{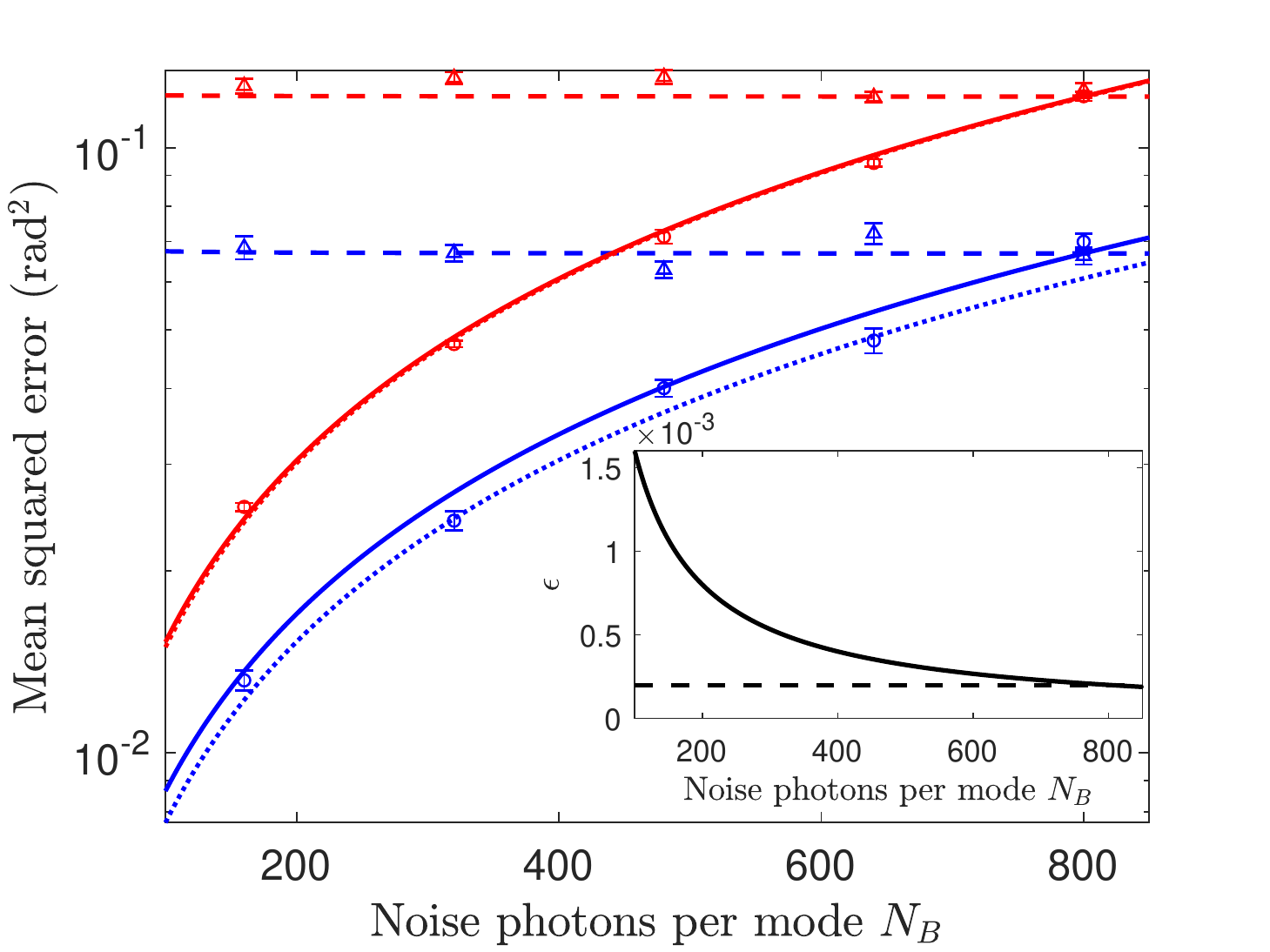}
    \caption{MSE errors vs background noise levels for entanglement-enhanced covert sensing (blue) and classical covert sensing (red). Experimental data (triangles) and theoretical model (dashed curves) in the fixed covertness regime; experimental data (circles) and theoretical model (solid curves) in the fixed probe power regime. Dotted lines: QCRBs in the fixed probe power regime. Inset: Covertness parameter vs noise photons per mode for the fixed covertness (dashed line) and fixed probe power (solid line) regimes. $\theta=\pi/2$,  $T=125$ $\mu$s, $\kappa$=0.0165. $N_S=8\times 10^{-4}$ for solid curves and QCRB ;$N_S/N_B=10^{-6}$ for dashed curves.}
    \label{fig:MSE}
\end{figure}

In light of Eq.~\ref{eq:epsilon}, one needs to increase the probe power at higher background noise levels to maintain a constant $N_S/N_B$ in the fixed $\epsilon$ regime. Choosing $\epsilon =2\times 10^{-4}$ over a sensing channel with transmissivity $\kappa_E=0.36$, the measured estimation MSEs (triangles) in Fig.~\ref{fig:MSE} show an expected constant behavior and an excellent agreement with the theoretical model (dashed lines). The estimation MSEs for the entanglement-enhanced covert sensing (blue) situate below those for classical covert sensing (red), thereby demonstrating a quantum advantage. The estimation MSEs (circles) in the fixed probe power regime also closely match the theoretical model (solid lines). Notably, the measured estimation MSEs approach the QCRBs (dotted curves), showing that both the entanglement-enhanced and classical covert sensing protocols are operating near their quantum optima. The corresponding covertness parameters in the two regimes are plotted in Fig.~\ref{fig:MSE} inset.

\begin{figure}[bth]
    \centering
    \includegraphics[width=0.53\textwidth]{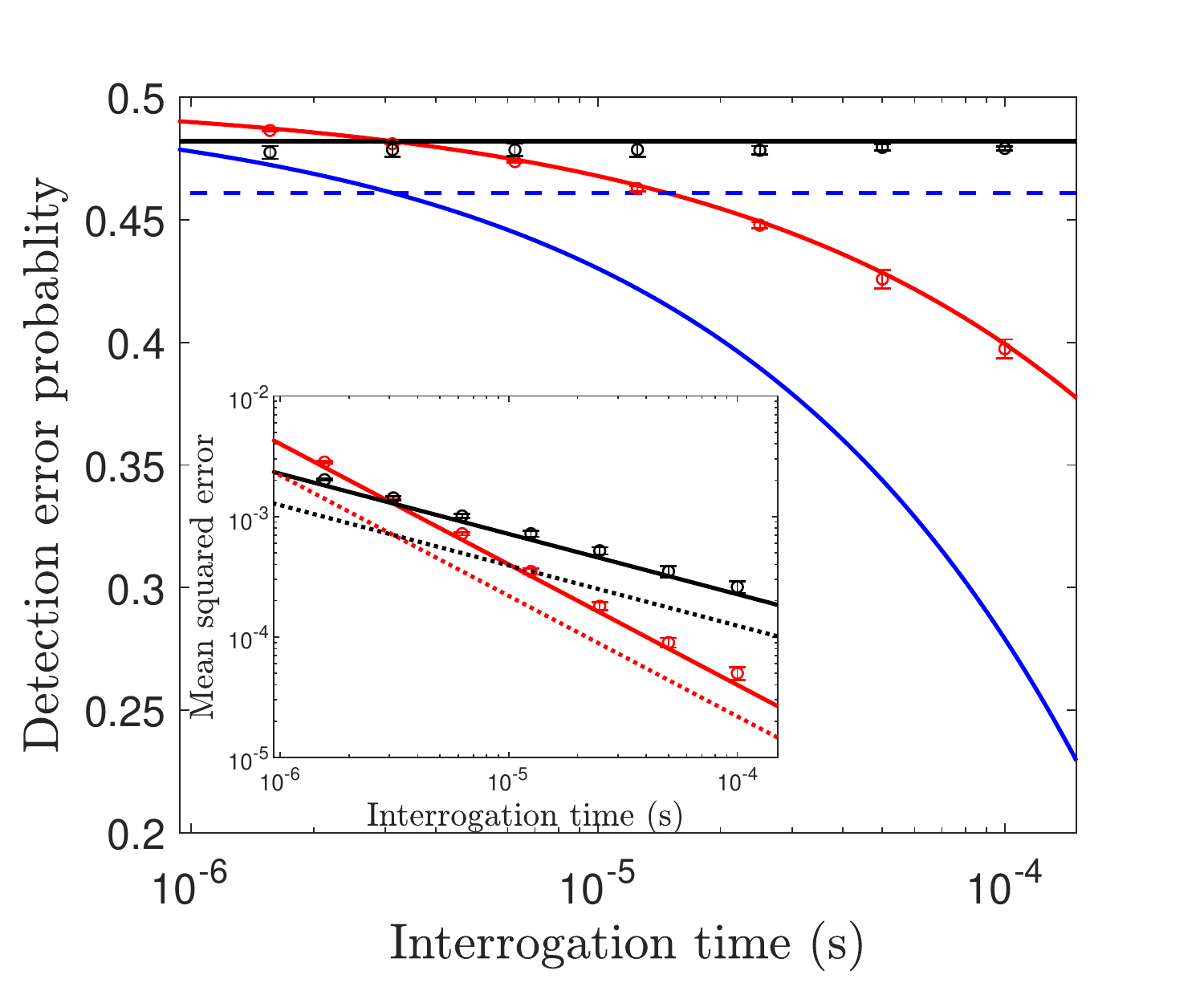}
    \caption{Test of Willie's detection error probability when the square-root law is obeyed (black) or violated (red). Thermal-loss sensing channel emulated by a 50:50 beam splitter ($\kappa_E=0.5$). $N_B=1280$. $\kappa N_S\sqrt{M}=200$ to obey the square-root law (black and blue dashed curve). $\kappa N_S/N_B=6.25\times 10^{-5}$ to violate the square-root law (red and blue curves). Dots: experimental data. Red and black curves: theory. Blue curves: lower bound for Willie's detection error probability. Inset: corresponding MSEs for classical (solid) and entanglement-enhanced (dotted) covert sensing around $\theta=\pi/2$, showing different scaling behaviors in obeying or violating the square-root law. MSE data for entanglement-enhanced covert sensing not taken due to limited photon flux at the source.}
    \label{fig:Covertness}
\end{figure}

Eq.~\ref{eq:epsilon} dictates that the per-mode probe photon number, $N_S$, needs to scale as $1/\sqrt{M}$ to maintain a constant covertness parameter at a given background noise level, leading to a square-root scaling for the signal-to-noise ratio with respect to the number of employed signal-idler mode pairs, viz. $MN_S/N_B \propto \sqrt{M}$, which is a signature for covert communication and sensing protocols~\cite{bash2015quantum,azadeh16quantumcovert-isit,bullock2019fundamental,gagatsos20codingcovcomm,bash12sqrtlawisit,bash2013limits,bash2015hiding,bash17qcovertsensingisit,goeckel17sensinglinsystems-asilomar,gagatsos2019covert}. We experimentally test the square-root law and report the result in Fig.~\ref{fig:Covertness}. Willie's detection error probabilities are measured at a range of $M$'s by varying the interrogation time $T$. Following the square-root law of $N_S \propto 1/\sqrt{M}$, Willie's detection error probabilities stay at a constant at a cost of a reduced slope for the MSE vs $T$ scaling, as illustrated by the experimental data (black dots) and the associated theoretical model (black curve) in Fig.~\ref{fig:Covertness}. In contrast, fixing probe power irrespective of the interrogation time violates the square-root law, resulting in an undesired reduction of Willie's detection error probabilities, as evidenced in the experimental data (red dots) and theory (red curve). The scaling of MSEs in obeying or violating the square-root law is illustrated in the inset Fig.~\ref{fig:Covertness}, unveiling a tradeoff between the measurement precision and covertness.

{\em Discussion.---}The PCR being optimum for entanglement-enhanced covert sensing but sub-optimum for quantum illumination~\cite{guha2009gaussian} unveils the fundamental disparity between two sensing regimes, parameter estimation and hypothesis testing. This situation is in analogy to phase estimation vs quantum-state discrimination based on coherent states: the HR is known to saturate the QCRB in estimating a phase shift imparted on a coherent state but fails to approach the ultimate Helstrom bound for discriminating two coherent states. Remarkably, the advantage of entanglement-enhanced covert communication protocols over their classical counterparts can scale as $1/\log(N_S)$~\cite{gagatsos20codingcovcomm,shi2020}, which diverges as $N_S \rightarrow 0$. This quantum advantage is in sharp contrast to the constant quantum advantage enabled by quantum illumination~\cite{tan2008quantum,zhang2015entanglement,shapiro2009quantum,lopaeva2013experimental,barzanjeh2015microwave}. Apart from being different from quantum illumination in the sensing regimes, entanglement-enhanced covert sensing bears a security constraint---the signal power and interrogation time need to be carefully chosen subject to the channel and covertness parameters.

Similar to covert communication, covert sensing hides the probe light in the noisy environment; however, the two tasks differ in their aims and figures of merit. Specifically, covert communication is evaluated by the number of bits that can be covertly transmitted while covert sensing concerns about the precision of parameter estimation. Both covert communication and sensing, be they based on classical or quantum resources, assume that the strong noise background is uncontrollable by Willie. Such a passive scenario would be well justified in the microwave domain where the natural blackbody radiation noise is abundant in the background. The blackbody radiation, however is negligible at optical wavelengths so that achieving covertness in optical communication or sensing needs to rest upon other effective background noise such as the sunlight or the internet traffic. Our present proof-of-concept covert sensing experiments are carried out at optical wavelengths but can be adapted to the microwave domain by leveraging efficient microwave-photonic transducers~\cite{jiang2020efficient}.

{\em Conclusions.---}We have demonstrated entanglement-enhanced covert sensing approaching the fundamental quantum limit set by the QCRB. The verified entanglement-enabled quantum advantage would pave a new route for quantum-enhanced secure sensing, communication, and information processing.

\section*{Acknowledgments}
This work is supported by Raytheon Technologies and in part by the Office of Naval Research (ONR) Grant No. N00014-19-1-2190 and National Science Foundation (NSF) Grants No. ECCS-1920742, CCF-1907918, CCF-2045530, and No. EEC-1941583. Q.Z. also acknowledges Defense Advanced Research Projects Agency (DARPA) under Young Faculty Award (YFA) Grant No. N660012014029. We thank Mark J. Meisner and Jaim Bucay for insightful discussions about the applications of the quantum technology.


%



\section*{Supplementary material}

# Supplemental Material
# Demonstration of Entanglement-Enhanced Covert Sensing

Shuhong Hao,[1] Haowei Shi,[2] Christos Gagatsos,[2] Mayank Mishra,[2] Boulat Bash,[3,2]
Ivan Djordjevic,[3,2] Saikat Guha,[2,3] Quntao Zhuang,[3,2] and Zheshen Zhang[1,3,2,∗]

[1]*Department of Materials Science and Engineering,
University of Arizona, Tucson, Arizona 85721, USA*
[2]*James C. Wyant College of Optical Sciences, University of Arizona, Tucson, Arizona 85721, USA*
[3]*Department of Electrical and Computer Engineering,
University of Arizona, Tucson, Arizona 85721, USA*

## I. EXPERIMENTAL DETAILS

### A. Experimental setup

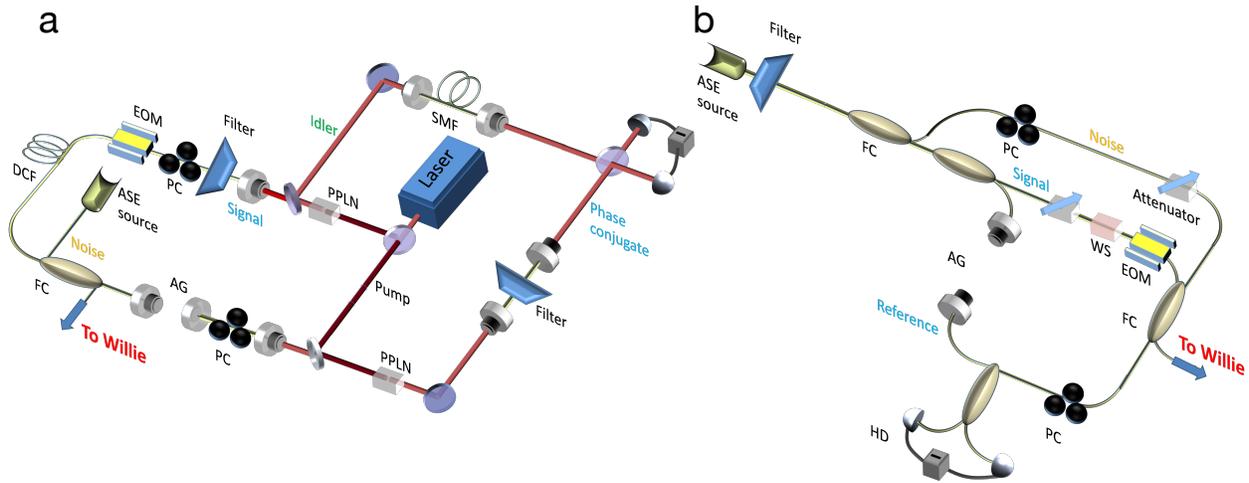

Figure S1. Detailed experimental setups for (a) entanglement-enhanced covert sensing and (b) classical covert sensing based on ASE light. AG: air gap. DCF: dispersion compensating fiber. EOM: electro-optic modulator. FC: fiber coupler. HD: homodyne detector. PPLN: periodically poled lithium niobate. SMF: single-mode fiber. WS: waveshaper. PC: polarization controller. ASE: amplified spontaneous emission source.

In the entanglement-enhanced covert sensing experiment, the entanglement transmitter consists of a temperature stabilized type-0 periodically poled lithium niobate (PPLN) crystal pumped by a continuous-wave (c.w.) 780-nm laser to produce non-degenerate signal and idler photon residing around 1590 nm and 1530 nm, respectively. A dichroic mirror (DM) separates the signal and idler photons, which are subsequently coupled into single-mode fibers (SMFs). A key to simultaneously achieve high covertness and estimation precision is a near-unity heralding efficiency of the idler photon conditioned on the paired signal photon such that all signal photons contribute to the phase sensing. To this end, the pump beam is loosely focused at the PPLN crystal to reduce the spontaneous parametric down conversion (SPDC) photons in higher-order spatial modes. Moreover, the collection optics for the signal and idler photons are designed, leading to a conditional heralding efficiency in excess of 99% [1]. A fiber-based optical filter confines the signal

photons to a 16-nm optical bandwidth, i.e., $W \sim 2$ THz, while the idler photons are retained in a low-loss SMF spool to be later used in the phase-conjugate receiver (PCR). The overall efficiency $\kappa_I$ for the idler distribution and storage is measured to be 96%. Due to the large optical bandwidths, dispersion on the signal and idler distorts their wavepackets and would reduce the efficiency of the PCR if not compensated. To maintain high storage efficiency for the idler, we leverage nonlocal dispersion cancellation [2–5] to *overcompensate* the signal using a spool of dispersion compensating fibers (DCFs) at the entanglement transmitter. An electro-optic modulator (EOM) driven by a programmable function generator serves as the interrogated phase object. The phase-shifted signal is then mixed with the thermal noise background produced by an amplified spontaneous emission (ASE) source on a fiber coupler, whose two output arms are diverted to Willie and the PCR. In the PCR, the signal is first delayed by a free-space air gap (AG) to fine tune and match with the optical path length of the idler and then combined with the pump on a DM. The pump and signal interact at a second PPLN crystal to generate the phase-conjugate light at 1530 nm. The phase-conjugate light is coupled into SMF via a collimator and then filtered to eliminate the residue signal photons. The filtered phase-conjugate light is coupled back to free space and interfere with the idler on a 50:50 beam splitter. The interference visibility is optimized to > 98%. The two output arms of the beam splitter are detected by a pair of photodiodes (Laser Components, InGaAs 1550) each with 99% quantum efficiency in a balanced configuration. The difference photocurrent is amplified by a transimpedance amplifier and then postprocessed to infer the probed phase.

In the classical covert sensing experiment, the output of an ASE source is split into three arms to serve as the signal, the reference, and the thermal background noise. The classical signal photons share the same photon statistics as those of the SPDC signal photons. Hence, Willie's capability of detecting the sensing attempt is independent of the type of the source and is fully determined by the brightenss of the signal, the magnitude of the background noise, and the transmissivity of the sensing channel. Two adjustable attenuators are used to fine tune the power of the signal and background noise to a desired level. To optimize the efficiency of interference between the signal and the reference, a waveshaper is employed to compensate for their differential dispersion. As in the entanglement-enhanced sensing experiment, an EOM applies the probed phase on the signal, and the background noise is then mixed in through a fiber coupler, whose two output arms goes to the classical receiver and Willie. At the classical receiver, the reference is first delayed in a free-space AG to match the propagation distance with the signal. The time-matched signal and reference are then mixed on a 50:50 fiber coupler followed by two photodetectors to perform a balanced homodyne measurement. The difference photocurrent is used to estimate the phase.

### B. Calibration

In our experiment, sensing is executed over $T$ seconds employing $M = WT$ signal-idler mode pairs. With a $P$-Watt beam, the number of photons per mode is $N_S = PT/Mh\upsilon = P/Wh\upsilon = P\lambda/Whc$, where $h = 6.626 \times 10^{-34} J/s$ is the Planck constant, $c \approx 3 \times 10^8$ m/s is the speed of light, the central wavelength in the experiment is $\lambda = 1550$ nm, and the optical bandwidth is $W = 2$ THz.

A photodetector converts the $P$-Watt beam in a coherent state to a photocurrent of $I = RP$, where $R$ is known as the responsivity of the detector. In our experiment, $R \approx 1.0$ A/W, corresponding to a quantum efficiency of 80%. Due to the random arrival time of photons, the photocurrent is intrinsically stochastic. The standard deviation of the photocurrent derived from either semi-classical or quantum photodetection theory reads

$$\Delta I_{\text{coh}} = \sqrt{2eI\Delta f}, \qquad (S1)$$

where $e = 1.6 \times 10^{-19} C$ is the electron charge and $\Delta f$ is the bandwidth of the photodetector set by, e.g., an electrical low-pass filter. Empirically, we choose $\Delta f = 0.75/T$ to optimize the sensing performance.

For a $M$-mode thermal light with $N_S$ photons per mode, the produced photocurrent is $I = \eta eWN_S$, where $\eta$ is the efficiency of the photodetector. The photon-number variance of the thermal state is $\langle \Delta^2 N \rangle = M \langle N_S \rangle + M \langle N_S^2 \rangle$, as compared to the photon-number variance $\langle \Delta^2 N \rangle = \langle N_S \rangle$ for the coherent state subject to the same per-mode photon number $N_S$. Therefore, the photon-number variance for the thermal state is $N_S + 1$ times of the photon-number variance for the coherent state. Consequently, the standard deviation for photocurrent of the thermal light reads

$$\Delta I_{\text{th}} = \Delta I_{\text{coh}} \sqrt{\eta N_S + 1} = \sqrt{2eI\Delta f + 2I^2\Delta f/W}. \qquad (S2)$$

The calibration uses a photodetector with a transimpedance gain of $\mathbb{G}_1 = 1.0 \times 10^4$ V/A (for 1 mega $\Omega$ load) to convert the photocurrent into a voltage signal that is measured in the frequency domain by an electrical spectrum analyzer (SRS SR760). A flat amplitude spectral density (ASD) in terms of $V_{\text{rms}}/\sqrt{\text{Hz}}$ is anticipated, derived as

$$\rho_{\Delta U_1} = \mathbb{G}_1 \sqrt{2ePR + 2(PR)^2/W}. \tag{S3}$$

In the experiment, $\mathbb{G}_1$ is calibrated by measuring the ratio of DC voltage and input light power. However, due to the intensity fluctuation of the ASE light, low-frequency (< 15 kHz) noise shows up in the direct detection spectrum, represented by the red curve in Fig. S2a. Such practical noise from the source would weaken Willie's capability of detecting the sensing attempt. To endow Willie full detection power, we exploit a balanced receiver (Thorlabs PDB450C) with transimpedance gain $\mathbb{G}_2 = 1.16 \times 10^5$ V/A (for 1 mega $\Omega$ load) to cancel the low-frequency noise. As a first step, we first test the performance of balancing by evenly splitting the ASE light and using the two photodetectors to take measurements. By optimizing the beam-splitter ratio, the polarization of the light at both photodetectors, and the optical loss, we are able to eliminate the intensity noise down to frequencies near the d.c. With $P = 80$ $\mu$W at each photodetector, the spectra of the voltage signals are depicted in Fig. S2b for the cases of imbalanced detection (red curve) and balanced detection (black curve). Evidently, the low-frequency noise is eliminated by appropriately balancing the optical power at the two photodetectors, leading to a shot-noise limited ASD of

$$\rho_{\Delta U_{\text{SNL}}} = 2\mathbb{G}_2 \sqrt{ePR}. \tag{S4}$$

To assess Willie's probability to detect the probe signal embedded in a pure thermal noise background, two orthogonal polarization modes of the light from the ASE source are measured by a pair of balanced photodetectors. In doing so, the joint low-frequency intensity-fluctuation noise of both polarization modes is cancelled while their independent thermal fluctuations are preserved, yielding an ASD of

$$\rho_{\Delta U_2} = 2\mathbb{G}_2 \sqrt{ePR + (PR)^2/W}, \tag{S5}$$

as depicted in the green curve of Fig. S2b.

The theoretical model for $\rho_{\Delta U_{\text{SNL}}}$, $\rho_{\Delta U_1}$, and $\rho_{\Delta U_2}$ are compared with experimental noise ASD data, shown in Fig. S2c with black, blue, and red curves, respectively. The transimpedance gain for the theoretical model and experimental data of $\rho_{\Delta U_1}$ is rescaled to that of $\rho_{\Delta U_{\text{SNL}}}$ and $\rho_{\Delta U_2}$ such that the three ASDs can be compared on an equal footing. The experimental data acquired at 50 kHz and 500 average times show excellent agreement with the theory model.

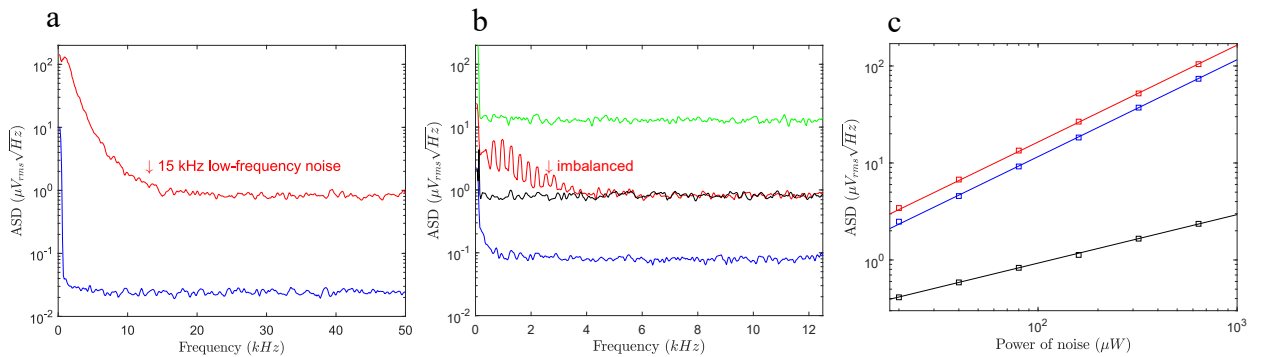

Figure S2. Experimental ASD data for (a) single-detector scheme using a photodiode (Thorlabs FGA01FC) and a current amplifier (Femto DLPCA-200) and (b) balanced-detection scheme (Thorlabs PDB450C). Data averaged over 40 average times. The optical power is 80 $\mu$W at each photodetector, yielding optical noise power much higher than the electrical noise floor (blue curve). (c) ASD data as a function of background noise power. Data averaged over 500 average times.

3

## C. Phase-locking module

In the classical covert sensing experiment, it is critical to stabilize the relative phase between the returned signal and the locally retained reference. Likewise, phase stability between the phase-conjugate signal and the idler is desired in the entanglement-enhanced covert sensing. To this end, we have developed a phase-locking module, described in Fig. S3, for both sensing scenarios. We use an EOM (Thorlabs LN65S in the entanglement-based experiment and Photonics iXblue MPZ-LN-10 in the classical experiment) to feedback the error signal to stabilize the phase, in addition to applying the probing phases as articulated at the outset.

In both the entanglement-enhanced and classical covert sensing setups, the output of the balanced receiver is split into two arms, one going to the oscilloscope to record data and the other connecting one input port of an electronic multiplier (AD835). The other input port of the multiplier takes a periodic 0/1 signal sharing the same frequency as the driving voltage that generates the phase for sensing. The output of the multiplier is further processed by a proportional–integral–derivative (SRS SIM960) controller to produce an error signal, which is then combined with the driving voltage on a summing amplifier (SRS SIM980) to drive the EOM. The multiplier periodically switches on the error signal to carry out phase locking. Phase sensing is then executed when phase locking is off.

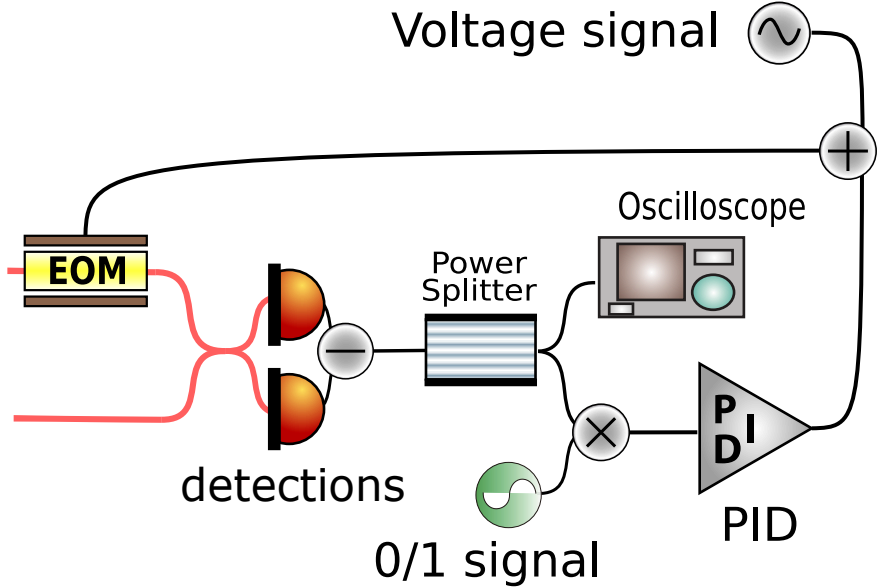

Figure S3. Schematic for the phase-locking module. EOM: electro-optic modulator. PID: proportional–integral–derivative controller.

The interleaved phase-locking and phase-sensing data are depicted in Fig. S4. First, we apply two different sensing phases and choose a probe signal and background noise power ratio of $P_B/P_S = 20000$ to test the phase-locking module. The results are shown in Fig. S4a-e. Fig. S4a and S4b illustrate, respectively, the phase-sensing driving voltage and the 0/1 signal that switches on/off phase locking. When phase locking is on, the relative phase between the signal and reference (the phase conjugate and idler) in classical covert sensing (entanglement-enhanced covert sensing) is stabilized to $\pi/2$. When phase locking is off, the phase-sensing driving voltage produces one of two optical phases around $\pi/2$. The output of the balanced receiver is filtered by a 500 kHz low-pass filter, and the results are plotted in Fig. S4d. As a comparison, the detector output for a free-running system without the phase-locking module is depicted in Fig. S4c. We next use a ramp driving voltage to generate the seven sensing phase, 0, $\pi/6$, $\pi/3$, $\pi/2$, $2\pi/3$, $5\pi/6$, $\pi$, as shown in Fig. S4e. By fitting the detector output to a cosine curve (Fig. S4f), we are able to infer a $V_\pi = 4.1$ V for Photonics iXblue MPZ-LN-10 used in classical covert sensing and $V_\pi = 8.6$ V for Thorlabs LN65S used in entanglement-enhanced covert sensing.

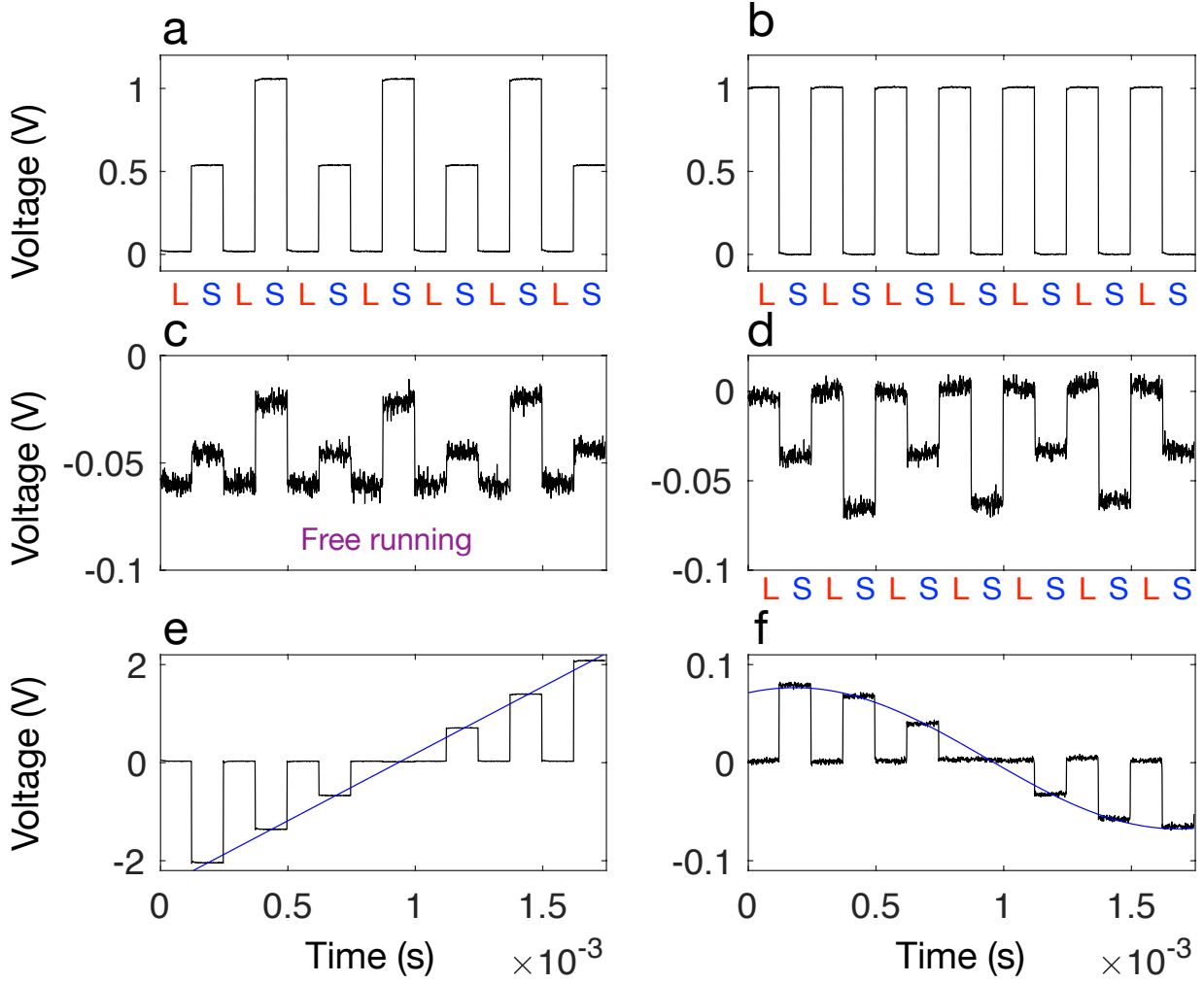

Figure S4. Phase locking results. (a) Driving voltage to create the binary phase pattern. (b) Input to the multiplier. (c) Free-running mode without phase locking. (d) Output of the balanced receiver in sensing the binary phases. Phase locking is on. (e) Driving voltage to create a phase ramp. (f) Output of the balanced receiver in sensing the phase ramp. Phase locking is on. Blue curve: a cosine signal fit to the output voltage. Interrogate time: $T = 125$ $\mu$s.

### D. Test of Willie's detection error probabilities

To test Willie's detection error probabilities, the probe signal and ASE noise feeds the two input ports of a 50:50 beam splitter. Willie captures the photons from one of the output port and either detects them directly using a single photodetector (Fig. S5a) or performs a balanced measurement (Fig. S5b) in conjunction with the orthogonal polarization of the ASE light from the source. The output of either detector is filtered by an analog filter (SRS SIM965 Butterworth type with a 48 dB/octave roll-off slope). An oscilloscope (LeCroy WavePro 604HD) records two traces of either detector at a sampling rate of $5 \times 10^7$ samples/s with the probe signal on or off. Willie's detection error probabilities are then inferred using the optimal decision threshold applied to both voltage traces.

We next develop a theoretical model for Willie's detection error probabilities. In the balanced-detection scheme, each photodetector receives a polarization mode with power $P_B$ from the ASE source. Using an electrical low-pass filter with a cutoff frequency of $\Delta f$ to reduce the out-of-band noise, the standard deviation of the voltage signal be

5

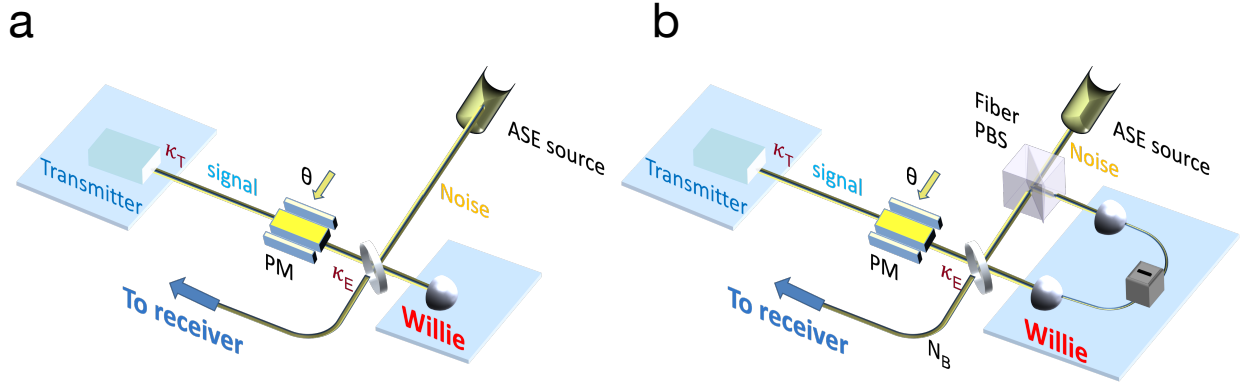

Figure S5. Willie's apparatus to detect the presence of the probe signal. (a) Single-detector scheme to perform a direct intensity measurement. (b) Balanced-detection scheme that eliminates the low-frequency intensity fluctuation noise. The sensing channel is modeled by a beam splitter reflecting $\kappa_E$ of the probe photons back to the receiver. The number of per-mode background noise photons is $N_B$ at the receiver.

derived using Eq. S5 as

$$\Delta U_2 = 2\mathbb{G}_2 P_B R \sqrt{\Delta f/W}, \tag{S6}$$

where the condition $N_B \gg 1$ has been applied to simplify the equation. Due to $M \gg 1$, the statistics of $\Delta U_2$ are effectively Gaussian. The presence of the probe signal with power $P_S$ increases the amplitude of the voltage signal from $U_{N_2}$ to $U_{S_2} = U_{N_2} + \mathbb{G}_2 P_S R$, so Willie's detection error probability is obtained by setting a decision threshold at $(U_{N_2} + U_{S_2})/2$:

$$P_{e_2} = \mathrm{erfc}\left[\frac{U_{S_2} - U_{N_2}}{2\sqrt{2}\Delta U_2}\right]/2 = \mathrm{erfc}\left[\frac{P_S}{4P_B\sqrt{2\Delta f/W}}\right]/2. \tag{S7}$$

In the experiment, $\Delta f = 0.75/T$ is chosen to optimize Willie's detection error probability, so Willie's detection error probability in the balanced-detection scheme is modeled as

$$P_{e_2} = \mathrm{erfc}\left[\sqrt{M/24}\frac{P_S}{P_B}\right]/2, \tag{S8}$$

which, shown in the red curve of Fig. S6, provides a precise model for the experimental data represented by the red dots in the same figure.

In the single-detector scheme, the standard deviation of the voltage signal can be similarly derived as

$$\Delta U_1 = \mathbb{G}_1 P_B R \sqrt{2\Delta f/W}, \tag{S9}$$

assuming a flat noise spectrum. However, the low-frequency noise precludes a first-principle model for the noise. To address this practical constraint, we introduce a fitting parameter $\gamma$ to Eq. S9's standard deviation model. Willie's detection error probability in the single-detector scheme then reads

$$P_{e_1} = \mathrm{erfc}\left[\sqrt{M/12\gamma}\frac{P_S}{P_B}\right]/2, \tag{S10}$$

which nicely fits the experimental data acquired from the single photodetector using $\gamma = 100$, as shown by the blue curve and dots in Fig. S6.

Due to the 50:50 beam splitter employed to emulate the sensing channel, $\kappa_E = 1 - \kappa_E = 0.5$, and Willie's noise

6

background equals to that at the receiver, so $P_S/P_B = \kappa N_S/N_B$. The lower bound for Willie's detection error probability is then derived as (see Sec. II E)

$$\mathbb{P}_e^{(w)} \geq 1/2 - \epsilon = 1/2 - \frac{\sqrt{M}\kappa N_S}{4N_B} = 1/2 - \frac{\sqrt{M}P_S}{4P_B}, \tag{S11}$$

as plotted in the green curve of Fig. S6.

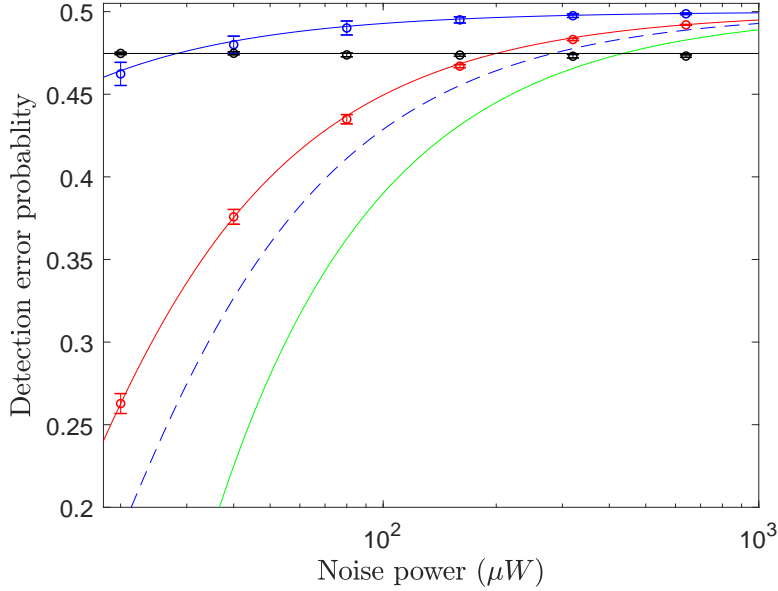

Figure S6. Evaluation of Willie's detection error probability in single-detector scheme (blue) and balanced-detection scheme (red). Blue dashed curve: theory for ideal single-detector scheme without being plagued by the low-frequency noise. Blue solid curve: fitted theory for practical single-detector scheme. Red curve: theory for balanced-detection scheme. Black curve: theory for balanced-detection scheme with fixed $P_B/P_S$. Green curve: QRE-based lower bound for Willie's error probability. $P_S = 20$ nW for red blue and green curves. $P_B/P_S = 10000$ for black curve. Dots: experimental data. Interrogation time $T = 2.42$ $\mu$s.

## II. THEORETICAL FRAMEWORK

### A. Model for phase sensing

The model for phase sensing through a lossy and noisy environment is shown in Fig. S7: the probe is modulated by a thermal loss phase shift channel $\mathcal{K}_\theta^{\kappa,N_B}$ characterized by an unknown phase $\theta$, a fixed transmissivity $\kappa$, and a fixed per-mode thermal noise number $N_B$.

For the entanglement-enhanced scenario, we consider a two-mode squeezed vacuum (TMSV) input state shared by the signal mode $\hat{a}_S$ and the idler mode $\hat{a}_I$. The state in number bases is written as

$$|\zeta\rangle = \sum_{n=0}^{\infty} \sqrt{\frac{N_S^n}{(N_S+1)^{n+1}}} |n\rangle_S |n\rangle_I, \tag{S12}$$

where $N_S$ is the average number of photons carried by the signal or the idler mode. After sensing over the channel $\mathcal{K}_\theta^{\kappa,N_B}$, each signal-idler pair $\{a_{R_m}, a_{I_m}\}$ conditioned on the phase $\theta$ is in a Gaussian state fully characterized by the

7

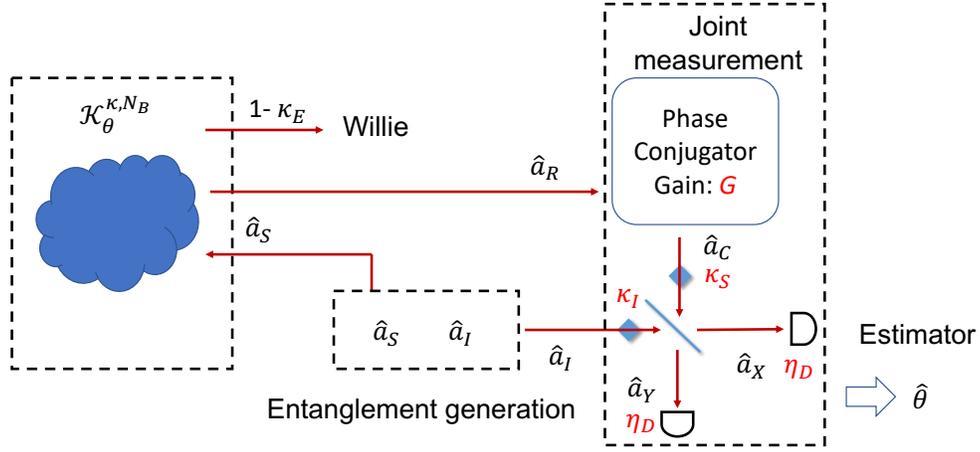

Figure S7. The schematic of the PCR, with experimental imperfections highlighted in red. $M$ i.i.d. probes are employed in sensing.

covariance matrix [6]

$$\mathbf{\Lambda}_\theta^{\text{TMSV}} = \begin{pmatrix} (2(N_B + \kappa N_S) + 1)\mathbf{I} & 2C_p \mathbf{R}_\theta \\ 2C_p \mathbf{R}_\theta & (2N_S + 1)\mathbf{I} \end{pmatrix}, \tag{S13}$$

where $\mathbf{R}_\theta = \text{Re}\left[\exp(i\theta)(\mathbf{Z} - i\mathbf{X})\right]$. Here $\mathbf{Z}$ and $\mathbf{X}$ are the $2 \times 2$ Pauli matrices. The amplitude of the cross correlation in each mode pair is $C_p = \sqrt{\kappa N_S (N_S + 1)}$.

For the classical sensing scenario, the ASE source is split into two arms to generate pairs of two-mode Gaussian states of $\hat{a}_S^{(l)}$ and $\hat{a}_R^{(l)}$ with mean photon number per mode being $N_S$ and $N_R$ respectively, where $1 \leq l \leq M$ is the mode-pair index. After $\hat{a}_S^{(l)}$'s propagating through the sensing channel $\mathcal{K}_\theta^{\kappa,N_B}$, the covariance matrix of each two-mode Gaussian state becomes

$$\mathbf{\Lambda}_\theta^{\text{ASE}} = \begin{pmatrix} (2(N_B + \kappa N_S) + 1)\mathbf{I} & 2C_p^{\text{ASE}} \mathbf{R}_{\theta,\text{ASE}} \\ 2C_p^{\text{ASE}} \mathbf{R}_{\theta,\text{ASE}}^\dagger & (2N_R + 1)\mathbf{I} \end{pmatrix} \tag{S14}$$

where $\mathbf{R}_{\theta,\text{ASE}} = \text{Re}[\exp(i\theta)(\mathbf{I} - \mathbf{Y})]$, and the amplitude of the cross correlation in each mode pair is $C_p^{\text{ASE}} = \sqrt{\kappa N_S N_R}$.

### B. Quantum Fisher information for noisy phase estimation

In single-parameter estimation, the quantum Fisher information can be obtained from the Uhlmann fidelity $\mathcal{F}(\rho, \sigma) = \text{tr}\left(\sqrt{\sqrt{\rho}\sigma\sqrt{\rho}}\right)^2$ as follows:

$$\mathcal{J}_\theta = \lim_{d\theta \to 0} 8 \frac{1 - \sqrt{\mathcal{F}(\rho_\theta, \rho_{\theta+d\theta})}}{d\theta^2}. \tag{S15}$$

Consider using the coherent state $\left|\sqrt{N_S}\right\rangle$ as the probe, the returned state at the receiver, $\mathcal{K}_\theta^{\kappa,N_B}(\left|\sqrt{N_S}\right\rangle\langle\sqrt{N_S}|)$, is a displaced thermal state with displacement $e^{i\theta}\sqrt{\kappa N_S}$ and a thermal noise contribution $N_B$. The fidelity can be derived

8

as [7]

$$\mathcal{F}^{\text{coh}}(\rho_\theta, \rho_{\theta+d\theta}) = \exp\left[-\frac{2\kappa N_S (1 - \cos(d\theta))}{1 + 2N_B}\right]. \tag{S16}$$

Thus the Fisher information can be obtained as

$$\mathcal{J}_\theta^{\text{coh}} = \frac{4\kappa N_S}{1 + 2N_B}. \tag{S17}$$

The variance of phase estimation is bounded since phase resides in $[0, 2\pi)$. For the Cramér-Rao bound to not diverge, the Fisher information is considered a measure of the sensing performance in an asymptotic manner where the $M$ repetitive measurements provide a factor $1/M$ that sufficiently reduces the variance below $2\pi$ such that the signal-to-noise ratio is high. Also, it is worthy to note that coherent state is the optimal input among classical states with a positive P-function, due to convexity of quantum Fisher information and the Fisher information of Eq. (S17) being linear in mean photon number $N_S$, i.e., $\mathcal{J}_\theta(\int dP_x |x\rangle\langle x|) \leq \int dP_x \mathcal{J}_\theta(|x\rangle\langle x|) = 4\kappa \int dP_x |x|^2/(1 + 2N_B) = \mathcal{J}_\theta^{\text{coh}}$.

With the covariance matrix Eq. (S14) of the returned signal-reference pair, the fidelity and thus the Fisher information are derived following the procedure outlined in Ref. [8]:

$$\mathcal{J}_\theta^{\text{ASE}} = \frac{4\kappa N_R N_S}{(1 + 2N_B)N_R + \kappa N_S + N_B}. \tag{S18}$$

which coincides with the quantum Fisher information of coherent states in the limit of $N_R \gg N_B, N_S$

$$\mathcal{J}_\theta^{\text{ASE}} \simeq \frac{4\kappa N_S}{1 + 2N_B} = \mathcal{J}_\theta^{\text{coh}}. \tag{S19}$$

Recently, an upper bound on the Fisher information is obtained [9]. While the full expression is lengthy and depends on the input state's mean photon number $N_S$ and the variance of photon number $\Delta_{N_S}^2$, one can show that the upper bound is maximized at $\Delta_{N_S}^2 \to \infty$, giving

$$\mathcal{J}_\theta^{\text{UB}} = \frac{4\kappa N_S \left(\kappa N_S + (1-\kappa) N_B' + 1\right)}{(1-\kappa)\left[\kappa N_S \left(2N_B' + 1\right) - \kappa N_B' \left(N_B' + 1\right) + \left(N_B' + 1\right)^2\right]}, \tag{S20}$$

where $N_B' = N_B/(1-\kappa)$ is the mean photon number of the thermal state at the environment mode in the Stinespring representation of the channel [9]. In the limit of $\kappa N_S \ll (1-\kappa)N_B$ and $(1-\kappa)N_B \gg 1$, The Fisher information upper bound becomes $\mathcal{J}_\theta^{\text{UB}} \simeq 4\kappa N_S/N_B(1-\kappa)$. Further assuming $\kappa \ll 1$, one has $\mathcal{J}_\theta^{\text{UB}} \simeq 4\kappa N_S/N_B$. This bound in fact covers the use of arbitrary entanglement between different probes, and between the probes and the ancillae.

Now we derive the quantum Fisher information for phase sensing based on the TMSV state. Using the methods from Ref. [8] to obtain the fidelity for the covariance matrix in Eq. (S13), the quantum Fisher information is derived as

$$\mathcal{J}_\theta^{\text{TMSV}} = \frac{4\kappa N_S (N_S + 1)}{1 + N_B (1 + 2N_S) + N_S (1 - \kappa)}. \tag{S21}$$

In the limit of $N_B \gg 1, \kappa \ll 1, N_S \ll 1$, we have

$$\mathcal{J}^{\text{UB}} \simeq \mathcal{J}_\theta^{\text{TMSV}} \simeq 2\mathcal{J}_\theta^{\text{coh}}. \tag{S22}$$

As such, entanglement gives rise to a factor of two advantage in the quantum Fisher information. Indeed, phase sensing based on the TMSV state is asymptotically optimal under our parameter setting. Because the upper bound also asymptotically holds for the multimode case, we conclude that the TMSV state is the general optimal state in the asymptotic regime $N_B \gg 1, \kappa \ll 1, N_S \ll 1$.

### C. Fisher information of PCR

We formulate a theoretical model based on Ref. [6] for the error analysis of phase sensing in the intermediate region where $G - 1$ is insufficient to enter the asymptotic regime described in the main text. In this region, higher-order terms in $N_S$ are no longer negligible. Our model accounts for experimental imperfections including the transmissivity after the phase conjugator ($\kappa_S$), the idler-storage efficiency ($\kappa_I$), and the detector quantum efficiency $\eta_D$, as shown in Fig. S7.

In what follows, we provide a full discussion about the imperfections and give a condition for the PCR to possess an advantage over the classical strategies using coherent or thermal states. Denote the returned signal-idler correlation of each probe as $\langle \hat{a}_R \hat{a}_I \rangle = C_p e^{i\theta}$. After phase conjugation, the receiver recombines the phase-conjugate and idler modes on a 50 : 50 beamsplitter and then detects the total photon difference between the two arms over the $M$ modes and calculate the total photon number difference $\hat{N} = \hat{N}_X - \hat{N}_Y$, which is Gaussian distributed in the limit of $M \gg 1$. Accounting for all the imperfections shown in Fig. S7, the mean $\mu(\theta)$ of $N$ conditioned on the phase $\theta$ can be calculated as

$$\mu(\theta) = M \cdot 2C_{CI}\eta_D \cos(\theta), \tag{S23}$$

and its associated variance is

$$\sigma^2(\theta) = M \cdot \left(\eta_D N_I + 2\eta_D^2 N_C N_I + \eta_D N_C + 2\eta_D^2 C_{CI}^2 \cos(2\theta)\right) \sim M \cdot \eta_D N_C, \tag{S24}$$

where $N_C = (G-1)\kappa_S(\kappa N_S + N_B + 1)$, $N_I = \kappa_I N_S$ and $C_{CI} = C_p \sqrt{(G-1)\kappa_I \kappa_S}$, with $C_p = \sqrt{\kappa N_S(1 + N_S)}$. For a Gaussian distribution, the classical Fisher information is given by $\mathcal{J}_\theta^{\text{PCR,M}}(\theta) = \int dx \, [\partial_\theta \ln(P_X(x|\theta))]^2 P_X(x|\theta) = [\partial_\theta \mu(\theta)]^2 / [\sigma^2(\theta)/M]$. Explicitly, the exact formula is

$$\mathcal{J}_\theta^{\text{PCR,M}} = M \cdot \frac{4\eta_D^2(G-1)\kappa\kappa_I\kappa_S N_S(N_S+1)\sin^2\theta}{\eta_D(N_I + N_C) + \eta_D^2\left(2N_C N_I + 2C_{CI}^2 \cos(2\theta)\right)}. \tag{S25}$$

At $N_B \gg 1$ and $N_S \ll N_B$, we use the approximation $N_C \simeq (G-1)\kappa_S N_B$ to get

$$\begin{aligned}
\mathcal{J}_\theta^{\text{PCR,M}} &\simeq M \cdot \frac{4\eta_D^2(G-1)\kappa\kappa_I\kappa_S N_S(N_S+1)\sin^2\theta}{\eta_D[(G-1)\kappa_S N_B + \kappa_I N_S] + 2\eta_D^2(G-1)\kappa_I N_S \kappa_S N_B} \\
&= M \cdot \frac{4\eta_D \kappa_I \kappa N_S(N_S+1)\sin^2\theta}{N_B(1 + 2\eta_D \kappa_I N_S) + \kappa_I N_S/[(G-1)\kappa_S]}.
\end{aligned} \tag{S26}$$

The first-order term $\kappa_I N_S / [(G-1)\kappa_S]$ in the denominator will be negligible when

$$(G-1)\kappa_S N_B \gg \kappa_I N_S. \tag{S27}$$

In this case, an entanglement-enabled quantum advantage in covert sensing is guaranteed. In the experiment $\kappa_S = 0.36$ accounts for the propagation loss in free space after the phase conjugator, collection efficiency of a collimator, and the transmissivities of two optical filters. $N_B/N_S$ ranges from $2 \times 10^5$ to $10^6$, $G - 1 = 0.257 \times 10^{-3}$, and $\kappa_I = 0.96$. As such, Eq. (S27) is fully justified to warrant a quantum advantage.

Now we consider the asymptotic behavior assuming ideal detectors and idler storage, i.e., $\eta_D \to 1, \kappa_I \to 1$. Discarding the next order in the variance, the asymptotic Fisher information reads

$$\mathcal{J}_\theta^{\text{PCR,M}} \sim M \mathcal{J}_\theta^{\text{TMSV}} \sin^2(\theta) \tag{S28}$$

in the limit of $N_B \gg 1, \kappa \ll 1, N_S \ll 1$. Thereby, at the point of $\theta = \pi/2$ PCR is the optimal measurement.

## D. Fisher information of homodyne receiver

In the homodyne receiver for the ASE source, we model the imperfection at the reference arm $\hat{a}_L$ by a pure loss channel with transmissivity $\eta_L$, while the signal interface and the detectors at the receiver side are assumed ideal. A 50:50 beam splitter recombines the signal and reference modes and produces modes $\hat{a}_X^{(l)} = \left(\hat{a}_S'^{(l)} + \hat{a}_L'^{(l)}\right)/\sqrt{2}$, $\hat{a}_Y^{(l)} = \left(\hat{a}_L'^{(l)} - \hat{a}_S'^{(l)}\right)/\sqrt{2}$ at its output ports. Then, a pair of balanced photodetectors measure the difference photon number summed over *all* $M$ modes, yielding $N \equiv N_X - N_Y$ with $N_X, N_Y$ being the classical random-variable outcomes of the photon-counting measurements $\sum_{l=1}^{M}\left(\hat{a}_X^{\dagger(l)}\hat{a}_X^{(l)}\right)$, $\sum_{l=1}^{n}\left(\hat{a}_Y^{\dagger(l)}\hat{a}_Y^{(l)}\right)$. Here, $N$ is a Gaussian random variable with mean $\mu(\theta) = 2M\sqrt{\eta_L \kappa N_L N_S}\cos(\theta)$ and variance $\sigma^2(\theta) = M[N_B + \eta_L N_L + 2\eta_L N_B N_L + \kappa N_S + 2\eta_L \kappa N_L N_S + 2\eta_L \kappa N_L N_S \cos(2\theta)]$. For a Gaussian distribution, the classical Fisher information is given by

$$\mathcal{J}_\theta^{HR,M} = [\partial_\theta \mu(\theta)]^2 / [\sigma^2(\theta)/M] = M \cdot \frac{2\eta_L N_L \kappa N_S \sin^2\theta}{N_B + \kappa N_S + 2\eta_L N_L [\kappa N_S(1 + \cos(2\theta) + N_B + 1/2]}. \tag{S29}$$

At the limit $\eta_L N_L \gg N_B$, $N_S \ll 1$,

$$\mathcal{J}_\theta^{HR,M} \simeq M \cdot \frac{2\kappa N_S \sin^2\theta}{N_B + 1/2} \sim M\mathcal{J}_\theta^{coh} \sin^2\theta. \tag{S30}$$

Therefore, the homodyne receiver achieves the optimal performance in classical phase sensing at the point of $\theta = \pi/2$.

## E. Covertness parameter

In both the entanglement-enhanced and classical covert sensing, Willie's marginal quantum state with (without) the probe signal is $\hat{\rho}_1^M$ ($\hat{\rho}_0^M$) with $N_1$ ($N_0$) photons per mode.

The covertness parameter is obtained via the relative entropy $D(\hat{\rho}_0^{\otimes M} \| \hat{\rho}_1^{\otimes M})$, whose calculation can be found in Refs. [6, 10, 11]. Using the additivity of relative entropy and thermal state properties (alternatively using Theorem 7 in Ref. [12]),

$$\begin{aligned} D(\hat{\rho}_0^{\otimes M} \| \hat{\rho}_1^{\otimes M}) &= MD(\hat{\rho}_0 \| \hat{\rho}_1) \\ &= M\left\{\log_2\left[\frac{N_1 + 1}{N_0 + 1}\right] + N_0 \log_2\left[\frac{N_0(N_1 + 1)}{N_1(N_0 + 1)}\right]\right\} \end{aligned} \tag{S31}$$

$$= \frac{M(N_1/N_0 - 1)^2}{2\ln(2)} + O\left(M(N_1/N_0 - 1)^3\right). \tag{S32}$$

Under the requirement of $P_E \geq 1/2 - \epsilon$, one can choose the relative entropy $D(\hat{\rho}_0^{\otimes M} \| \hat{\rho}_1^{\otimes M}) \leq 8\epsilon^2/\ln(2)$, so

$$\epsilon \simeq \frac{\sqrt{M}(N_1/N_0 - 1)}{4}. \tag{S33}$$

The sensing environment is a thermal-loss channel modeled as a beam splitter that reflects $\kappa_E$ portion of the probe signal back to the receiver. The background noise with $N_B/(1 - \kappa_E)$ photons per mode is injected through the other input port. Willie captures all photons from the beam splitter output port that does not connect to the quantum receiver. Straightforward calculations give $N_0 = \kappa_E N_B/(1 - \kappa_E)$ photons per mode in the absence of the probe signal and $N_1 = (1 - \kappa_E)\kappa_T N_S + \kappa_E N_B/(1 - \kappa_E)$ in the presence of the probe signal. Hence,

$$\epsilon = \frac{\sqrt{M}(1 - \kappa_E)^2 \kappa_T N_S}{4\kappa_E N_B} \propto \frac{\sqrt{M} N_S}{N_B} \tag{S34}$$



\end{document}